\documentclass[showpacs,aps,prb,twocolumn]{revtex4}
\usepackage{epsfig}

\begin{document}

\title{Conductance scaling at the band center of
wide wires with pure non--diagonal disorder}

\author{J. A. Verg\'es}
\email{jav@icmm.csic.es}

\affiliation{Instituto de Ciencia de Materiales de Madrid,
Consejo Superior de Investigaciones Cient\'{\i}ficas,\\
Cantoblanco, E-28049 Madrid, Spain.}

\date{February 2nd, 2001}

\begin{abstract}
Kubo formula is used to get the scaling behavior of the static
conductance distribution of wide wires showing pure non--diagonal disorder.
Following recent works that point to unusual phenomena
in some circumstances, scaling at the band center of wires of odd widths
has been numerically investigated. 
While the mean conductance shows a decrease that is only proportional
to the inverse square root of the wire length,
the median of the distribution exponentially decreases as a function of
the square root of the length.
Actually, the whole distribution decays as the inverse square root of the
length except close to $G=0$ where the distribution accumulates the weight
lost at larger conductances. It accurately follows the theoretical
prediction once the free parameter is correctly fitted.
Moreover, when the number of channels equals the wire length
but contacts are kept finite, the conductance distribution is still
described by the previous model. It is shown that the common origin
of this behavior
is a simple Gaussian statistics followed by the logarithm of the
$E=0$ wavefunction weight ratio of a system showing chiral
symmetry. A finite value of the two--dimensional mean conductance
is obtained in the infinite size limit.
Both conductance and the wavefunction statistics distributions
are given in this limit.
This results are consistent with the {\it critical} character of the $E=0$
wavefunction predicted in the literature. 
\end{abstract}


\maketitle

\section{Introduction}

The existence of peculiar properties that differentiate pure non--diagonal
disorder models from disorder models including diagonal disorder has a long
history. Probably, the first contribution along this line corresponds
to Dyson's work on a one--dimensional phonon model published in
1953\cite{dyson}. The existence of a divergent density of states
at the band center of such disordered systems implying a divergent
localization length was pointed in later works\cite{1d_a,1d_b}.
Since then many works on this subject have been published
(a representative list can be found in Ref.(\onlinecite{mudry0})).
The existence of some kind of delocalization transition at the
band center of such models certainly disturbs the
widely accepted statement saying that the specific form of
disorder does not matter in single parameter scaling theory\cite{economou}.

A bipartite lattice is a lattice that can
be divided into two sublattices such that the Hamiltonian changes sign
under a transformation that changes the sign of the wavefunction on one
sublattice. When pure non--diagonal disorder is considered
on a bipartite lattice, the electron-hole symmetry of the spectrum
is not destroyed by disorder.
This property has important consequences as, for example,
the existence of an eigenstate at $E=0$ for any disorder realization
of a system constituted by an odd number of sites
(the spectrum shows $\pm E_\alpha$ pairs plus a state at $E=0$).
Recently,
several works pointing to the exotic behavior of transport properties
of quantum wires showing chiral symmetry
have been published\cite{mudry0,mudry}. 
For example, scaling of the conductance strongly depends on the
parity of the number of channels along the wire.
Also, related activity in field theory has produced several models
demonstrating a delocalization transition in the vicinity of the
zero energy state\cite{campos}.

In this paper, the scaling properties of the simplest disorder model
preserving chiral symmetry have been carefully analyzed at the band center
of quantum wires. Numerical simulation has been used to get the whole
distribution of conductances. While the mean value of the conductance
decays algebraically for wires of odd width, an alternative measure of the
central value of distributions -the median- shows an exponential decay.
Actually, the larger part of the conductance measurements has an exponentially
small value for large wires. Therefore, although the conductance distribution
is certainly peculiar at the band center, I would still use the term
exponential localization when referring to the scaling behavior of the
conductance of long wires.
Results change when width and length of the wire coincide. In this case,
the numerical simulation presented in this paper uses a dot
geometry point of view keeping the size of the contacts finite
while dot area scaling proceeds. The conductance
distribution converges to a well defined limit that is compatible
with the predicted critical behavior of the state at the band center. 
It is shown that the analytic form used to fit conductance distributions
comes from a new underlying wavefunction statistics describing the
distribution of weight ratios of the $E=0$ wavefunction.

The format of the paper is as follows. Section II defines a quite
simple chain model that allows some analytical results and an unbound
numerical simulation. Section III gives a more general disorder model on
finite rectangular clusters of the square lattice. The way in which the
conductance is calculated is presented in Section IV.
Numerical results are given in Section V, first for wide disordered wires
and, second, for square clusters.
The last Section of the paper compiles the main conclusions
reached by this numerical study of conductance scaling.

\section{Toy model}

Let me begin with a detailed description of the scaling properties of the
conductance of a chain showing pure non--diagonal disorder. In this case,
sign changes of the hopping parameters do not matter and changes in their
absolute value should be considered. One case that allows some analytic
results consists of a chain with hopping parameters that randomly take
values of $1$ and $\sqrt{2}$ with equal probability.
Starting from the selfenergy of an ideal chain at its band center
$\Sigma_0=-i$,
successive selfenergies can be obtained through the disordered part of the
chain by means of the usual iterative sequence (see, for example,
Eqs. (7)-(10) in Ref.(\onlinecite{1d_b})):

\begin{equation}
\Sigma_{n+1}=-{{t_n}^2 \over \Sigma_n} \qquad,
\label{chain}
\end{equation}

\noindent
where ${t_n}^2$ takes values 1 or 2 with equal probability. Since
selfenergy remains purely imaginary, a more convenient form is possible:

\begin{equation}
-\Sigma_{n+1}/i={{t_n}^2 \over (-\Sigma_n/i)} \qquad.
\label{chain2}
\end{equation}

\noindent
The conductance $G$ of a sample is obtained from the selfenergy at the end
of the disordered chain part $\Sigma_N/i$:

\begin{equation}
G={{4 (-\Sigma_0/i) (-\Sigma_N/i)} \over
  [1+(-\Sigma_0/i)(-\Sigma_N/i)]^2} =
  {{4 (-\Sigma_N/i)} \over [1+(-\Sigma_N/i)]^2} \qquad.
\label{chain3}
\end{equation}

\noindent
Eq. (\ref{chain3}) shows
that the conductance varies between 0 and 1 as it corresponds to
a one--channel system. Actually, the conductance takes the value 1 only if
$-\Sigma_N/i=-\Sigma_0/i=1$.
Repeated use of Eq.(\ref{chain2}) shows that the form in which
the random hopping elements appear in the selfenergy expression is:

\begin{equation}
\frac
{{t_0}^2 {t_2}^2 {t_4}^2 ...} {{t_1}^2 {t_3}^2 {t_5}^2 ...} \qquad.
\label{chain4}
\end{equation}

\noindent
Therefore,
a perfect transmission through the chain is obtained when the number of
normal (hopping equal to 1) and strong (hopping equal to $\sqrt{2}$) bonds
of the numerator coincides with the corresponding numbers of the denominator.
Solving this simple combinatorial problem, one gets a probability 

\begin{equation}
p(G=1)=
{1 \over 2^N}
\sum_{n=0}^{N/2} \:
\bigg[
{(N/2)! \over {(N/2-n)! n!}}
\bigg]^2
\label{probability}
\end{equation}

\noindent
for the peak of the probability distribution of the conductance
at $G=1$\cite{curiosidad}.
Notice that the distribution is a sum of delta functions since the hopping
takes just two different values; for example, the peak below
the one at $G=1$ appears at $G=8/9$ and corresponds to
$-\Sigma_N/i=2$ or $-\Sigma_N/i=1/2$ (one extra strong bond either in the
numerator or the denominator).
An asymptotic expansion for large enough chain lengths can be found for
the sum in (\ref{probability}):

\begin{equation}
p(G=1) \sim \sqrt{2 \over {\pi N}} \qquad .
\label{law}
\end{equation}

\noindent
The scaling behavior of this peak is enough to explain an inverse
square root law for the {\it mean value} of $G$ given that other peaks
of the distribution decay in a similar or faster way with the size of the chain.
This is an interesting behavior since it perfectly coincides with scaling
predictions for wide wires of odd number of transversal modes\cite{mudry}.

\begin{figure}
\epsfig{figure=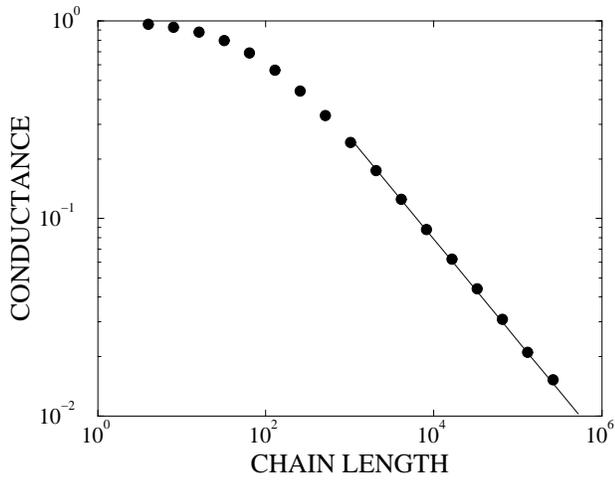,height=7cm,width=9cm}
\caption{Length scaling of the conductance of a disordered chain at $E=0$.
Average conductance is given in units of the conductance quantum
${e^2} / h$.
Although the variance of the conductance distribution is large,
the values given for the averages show error bars that
are sensibly smaller than the symbols representing the mean values.
The fit at large lengths by a power law curve of exponent
$-0.51 \pm 0.01$ is also shown.}
\label{toy.ps}
\end{figure}

\begin{figure}
\epsfig{figure=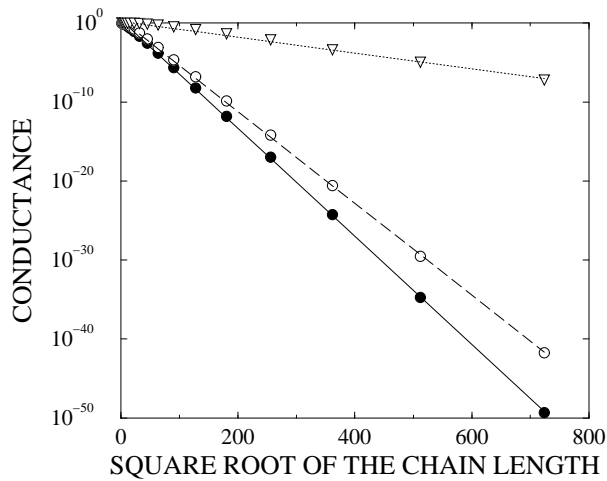,height=7cm,width=9cm}
\caption{Length scaling for the model of Fig. \ref{toy.ps} when the conductance
distribution is described by an alternative {\it central value} such as the
average value of the logarithm (filled circles), the median (empty circles)
or the conductance value at which the integral of the distribution reaches
a value of 0.9 (down triangles). All of them are fitted in the whole
range of chain lengths by an exponential law decaying as a function of the
square root of the sample length.}
\label{toy2.ps}
\end{figure}

Numerical simulations can be done as precise as necessary for this
simple model. This fact allows a detailed comparison with theoretical
predictions.
Fig. \ref{toy.ps} shows the scaling of the conductance as described by its
mean value. Root mean square deviation is about some tenths of the conductance
unit. In any case, the average conductance at large sizes ($N>1000$)
shows a power law scaling
with the distance that is compatible with the inverse square root law given
by Eq.(\ref{law}). According with this result, a disordered chain with
pure non--diagonal disorder shows non--standard scaling at the band center.
Nevertheless, alternative measures of the central value of the distribution
restore to some way the usual exponential decay of one--dimensional
conductances, {\it even} in the presence of chiral symmetry.
For example, the geometric mean\footnote{
The geometric mean is obtained from
the average of the logarithm of the random variable by:
$$<x>_{\rm geom}=\exp{<\ln(x)>}$$}
of the same conductance distributions
shows a much more pronounced decrease with length
(see Fig. \ref{toy2.ps}) although the corresponding standard
deviation is of the same order as the mean making doubtful its statistical
relevance.
But there are other alternatives which do give a good description of the
overall scaling of the distribution. Both the median\footnote{
The median of a probability distribution function $p(x)$ is the value
$x_{\rm med}$ for which larger and smaller values of $x$ are equally
probable:
$$
\int_{-\infty}^{x_{\rm med}} p(x) dx =\frac{1}{2}=
\int_{x_{\rm med}}^{\infty} p(x) dx
$$ }
or any definition based
on the value of the integral of the distribution between 0 and an arbitrary
upper limit $G_{\rm max}$ flow to exponentially small values.
The physical meaning is clear in this case:
half or more of the measures are exponentially small at large chain lengths.
Actually, the precise scaling law for these central value alternatives is:

\begin{equation}
G_0 \sim \exp \Big( - \sqrt{L / \xi} \Big) \qquad ,
\label{law2}
\end{equation}

\noindent
where $\xi$ gives a measure of the exponential localization length.
Fig. \ref{toy2.ps} shows that fits according to this law are
excellent over the whole length range.
The ultimate reason for such statistically disappointing
results is simply the unusual size scaling of the distribution.
While the major part of the distribution below $G=1$ decays in an algebraic
form, the weight of the distribution accumulates in an exponentially
small region near $G=0$ (Fig. \ref{mudry_hist.ps} in Section V illustrates
graphically this behavior).
In this way, the upper part of the distribution
dominates the scaling behavior of mean, root mean square deviation, etc.,
while
accumulation at $G=0$ gives the behavior of central value definitions based
on the integral of the distribution. In my opinion, these last definitions
are better suited for characterizing the whole distribution than standard
averages. Ultimately, one should look at the precise experimental protocol
followed to get a value of the conductance before making predictions about
the result of the measurements.

\section{Non--diagonal disorder model}

The lattice Hamiltonian describing random hopping on a $L \times M$ cluster
of the square lattice is :

\begin{equation}
\hat H = {\sum_{<l l'>} t_{l l'} \hat c_{l}^{\dag} \hat c_{l'}} \qquad,
\label{RHS}
\end{equation}

\noindent
where $\hat c_{l}^{\dag}$ creates an electron on site $l$,
$l$ and $l'$ are nearest--neighbor sites,
and $t_{l l'}$ is the hopping energy from site $l$ to $l'$.
It takes values 1 and -1 with equal probability. Let me refer to this model
as the Random Hopping Sign (RHS) model. 
Obviously, the square lattice can be divided into two sublattices such that
atoms belonging to one of them hop only to sites belonging to the other
sublattice when described by Hamiltonian (\ref{RHS}). Therefore,
this Hamiltonian changes sign under a transformation that changes the sign
of the electron operators on one sublattice.
Consequently, the spectrum is symmetric relative to the band center at $E=0$
for any disorder realization, i.e., for any values of the random variables
of the model $\{ t_{l l'} \}$.

The model given by Eq.(\ref{RHS}) is probably the simplest
two dimensional model showing chiral symmetry.
Many changes can be done to this model preserving chirality. For example,
absolute values of the hopping can fluctuate in addition to their signs or
complex values of the hopping parameters can be considered if random
magnetic fluxes are simulated. The results described below are not
sensible to any of these modifications of the disorder model.
Numerical values are somewhat changed but trends remain exactly the same.

\section{Conductance calculation}

Randomly generated samples of $L \times M$ clusters are connected
to ideal leads of width $M$. Typically,
$L \gg M$ in a wire geometry. The conductance of the whole system
is obtained using Kubo formula\cite{kubo}
within exact one-electron linear response theory.
Computational details are given elsewhere\cite{jav2,cpc}.
Let me just mention that the inversion of the Hamiltonian matrix
needed to get the Green function of the wire cannot proceed slab by slab
as it is usually done within an optimized code.
Numerical divergences take place owing to
the existence of a true eigenstate at exactly $E=0$
for any piece of the system showing an odd number of sites
(the matrix $E - \hat H$ is locally singular). Nevertheless,
numerical calculation proceeds straightforward when pivoting over the
whole Hamiltonian matrix that includes the ideal leads is allowed.

\section{Results}

\subsection{Wide disordered wires}

\begin{figure}
\epsfig{figure=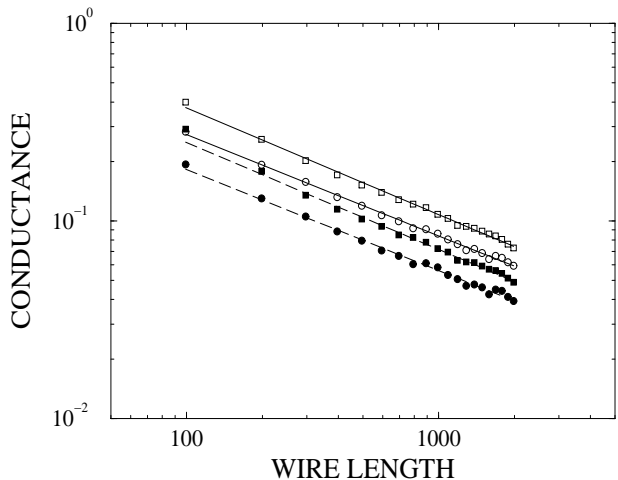,height=7cm,width=9cm}
\caption{Length scaling of the mean conductance of a wire of 9 (empty circles)
or 19 channels (empty squares) at the band center ($E=0$).
Average conductances are given in units of the conductance quantum
 ${e^2} / h$ and fitted to power law curves of exponents
$-0.51 \pm 0.01$ and $-0.54 \pm 0.01$, respectively.
Filled symbols give the values of the variance obtained for the
same sets of randomly generated samples. Dashed lines are just
2/3 of the previous fits.
Error bars are of the order of the symbols representing the averages.}
\label{mudry1.ps}
\end{figure}

The first aim has been the recovery of some important results obtained for
wide wires by Mudry, Brouwer, and Furusaki\cite{mudry}. In particular,
the exotic dependence of the scaling law on the parity of the
wire width is obtained for the present model (a simplification of their
Random Flux Model).
Open boundary conditions have been used to get conductances of
stripes of fixed width and number of open channels
(the wire width is equal to the number of channels
at the band center).
Sample lengths have been varied from 99 to 1980 in steps of 99.
As many as $10^4$ samples are necessary to get good values of
means and other central values of conductance distributions.
Fig. \ref{mudry1.ps} shows the scaling law for two typical
odd widths (9 and 19 channels) in a log-log plot.
Error bars are comparable to the symbols representing
the conductance averages. A power law fit to the numerical data is compatible
with a mean conductance scaling proportional to the inverse
square root of the sample length $L$:
$$
<G> \sim {1 \over {\sqrt{L}}}  .
$$
This is precisely the scaling law obtained in Ref.(\onlinecite{mudry}) for 
quantum wires of an odd section.
Notice that scaling proceeds smoothly without distinguishing odd and even
wire lengths.
As noted by these authors, the variance of the conductance distribution
is as large as its mean, making the mean a poor characteristic of the whole
distribution. Actually, the predicted relationship
$$
{<G^2> \over <G>} = \frac{2}{3}
$$
is accurately reproduced by the numerical data (see filled symbols of
Fig. \ref{mudry1.ps} and the dashed lines that are simply $2/3$ of the
previous fits). On the other hand,
when an even number of channels is studied, exponential scaling of the
average conductance together with an exponentially small typical deviation
is obtained in complete agreement with previous work\cite{mudry}.

\begin{figure}
\epsfig{figure=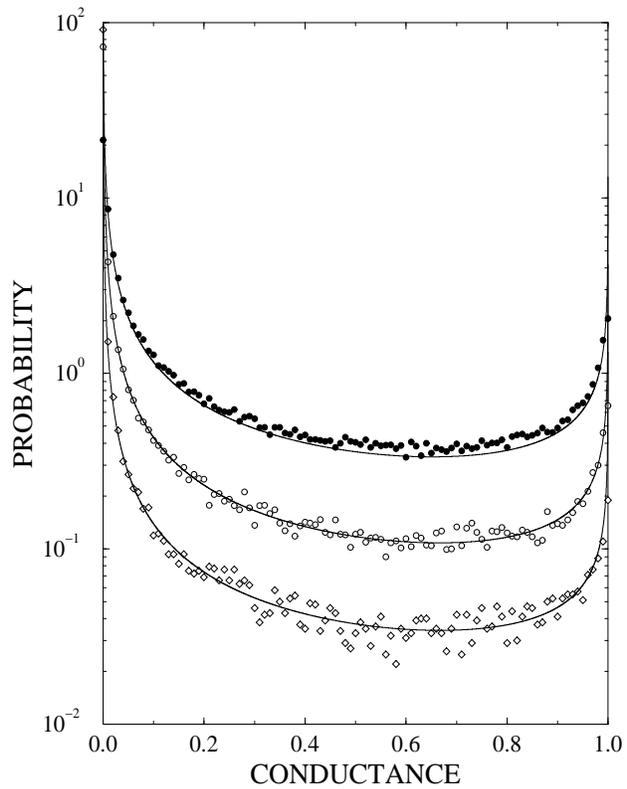,height=12.6cm,width=9cm}
\caption{Length scaling of the probability distribution of conductances
of a nine channel wire. Filled circles, empty circles and diamonds give
histograms for the statistics of $10^5$ samples of lengths 100, 1000 and
10000, respectively. Continuous lines give results obtained from
Eq. (\ref{p_law}). Conductance unit is ${e^2} / h$.}
\label{mudry_hist.ps}
\end{figure}

A deeper understanding of scaling of the conductance mean value
can be obtained from the analysis
of the whole conductance distribution. Here, conductances for $10^5$
samples of width $9$ and lengths $10^2$, $10^3$, and $10^4$ have been
compiled and the corresponding histograms plotted in Fig. \ref{mudry_hist.ps}.
It is clear that the probability for large values of the conductance
diminishes as the length of the wire increases. Actually, since the plot
is semilogarithmic the roughly equal separation between filled and
empty circles for one side and between empty circles and diamonds for the
other side, implies a power law decrease of the probability.
The weight that the probability distribution loses for large conductance
values goes close to $G=0$.
I have checked that the divergence of the probability
at $G=1$ is proportional to $1 / \sqrt{1-G}$ while the
divergence at the origin looks also algebraic but with an exponent
starting close to $-{ \frac{1}{2}}$ for small disorder and
decreasing towards $-1$ as disorder increases.

Although the conductance corresponding to ideal (non disordered)
wires equals their widths (9 and 19, in this case),
Fig. \ref{mudry_hist.ps} shows that conductance just reaches
the value of 1 for disordered systems.
It seems that just one channel is effective.
Therefore, it is tempting to compare these conductance distributions
with the one corresponding to one random channel within the orthogonal
universality class\cite{coe} arguing that before localization the
effect of disorder is just randomizing transport coefficients.
But the comparison is very bad since apart for
small differences all theories show probabilities that continuously
decrease from $G=0$ to $G=1$. In particular, the simplest result that
applies to a random channel described by scattering matrices of the
Circular Orthogonal Ensemble is:

\begin{equation}
p(G)=\frac {1} {2 \sqrt{G}}  \qquad,
\end{equation}

\noindent
that is quite different from the one obtained by numerical simulation.
There is a direct mathematical reason explaining this failure.
Typically, several states contribute to the Green function calculated
at an arbitrary energy within the spectrum of a disordered system.
But this is not the case when the Green function of a chiral system
is calculated at $E=0$ which is an eigenenergy of the isolated
system with an odd number of sites. In this situation, both the
Green function of the isolated finite system and the one
corresponding to the extended system including the leads
(and related to the previous one by a Dyson equation)
are dominated by the pole at $E=0$, that is, are completely determined
by the $E=0$ eigenfunction.
In the next subsection, I will exploit this feature to analyze
the conductance distribution as a consequence of a precise wavefunction
statistics. Meanwhile, let us return to the theory of Ref. (\onlinecite{mudry})
to analyze the numerical results.

Mudry, Brouwer, and Furusaki give the following expression for
the conductance distribution of wires with an odd number of channels
in the localized regime (see Eq. (4.10) of the
second paper of Ref. (\onlinecite{mudry})):

\begin{equation}
p(G)=\sqrt{C \over \pi} { {\exp \{ -C [{\rm arcosh}(G^{-1/2})]^2 \} }
 \over {G \sqrt{1-G}}} \qquad,
\label{p_law}
\end{equation}

\noindent
where

\begin{equation}
C={l \over {4 L}} {{M^2+M-2} \over {M-1}} \qquad,
\label{cte}
\end{equation}

\noindent
with $l$ the mean free path, $L$ the wire length, and $M$ the wire width
(which coincides with the number of channels at the bandcenter).
A second scaling parameter $\eta$ that characterizes the disorder on a
microscopic scale does not appear because it vanishes for the random
flux model (RFM) and the model studied here (see Section III) is
just a special case of RFM model\cite{jav2}.
When $C$ is small enough ($C < 0.01$) the distribution is very well
described by a much simpler expression (error smaller than one percent
for $G \ne 0$):

\begin{equation}
p(G) \simeq \sqrt{C \over \pi} {1 \over {G \sqrt{1-G}}}
\end{equation}

\noindent
which allows the evaluation of the mean and the variance of the distribution:

$$
<G> \simeq 2 \sqrt{C \over \pi}
$$
$$
<G^2> = \frac{2}{3} <G> \qquad.
$$

\noindent
Using the explicit form of $C$ (Eq.(\ref{cte})) an expression is obtained
that can be used to fit the remaining parameter $l$ once an enough
number of widths and/or lengths have been studied:

\begin{equation}
{<G>}^2 L \simeq {l \over \pi} {{M^2+M-2} \over {M-1}} \qquad.
\label{teoria}
\end{equation}

\begin{figure}
\epsfig{figure=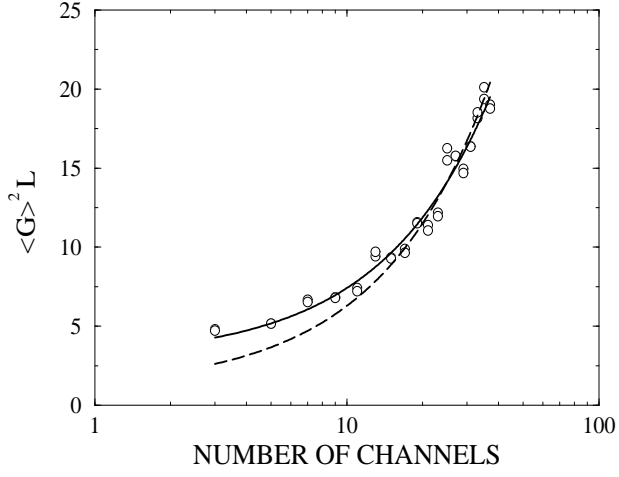,height=7cm,width=9cm}
\caption{Dependence of the mean conductance and variance
on the number of channels for a fixed length of 2000. The number of
samples used to get accurate values of the statistical values is $10^4$.
Two fits to the data are shown: one using Eqs. (4.11a) and (4.11b)
of the second paper in Ref.(\onlinecite{mudry}) with $l=1.646$ (dashed line)
and a linear fit ($C$ given by Eq.(\ref{lineal2})) (continuous line).}
\label{fit.ps}
\end{figure}

\noindent
Fig. (\ref{fit.ps}) shows a fit to the mean and variance values obtained for
sets of $10^4$ randomly generated
samples of length $L=2000$ and width $M$ from three to 37. While
Eq.(\ref{teoria}) works reasonably good for large widths it fails near
$M=3$. Actually, an alternative fit by a linear law works sensibly better:

\begin{equation}
{<G>}^2 L = a M + b \qquad.
\label{lineal}
\end{equation}

\noindent
Consequently,
the distribution given by Eq. (\ref{p_law}) will be used with
\begin{equation}
C = \frac{0.3515 M + 2.3050}{L}
\label{lineal2}
\end{equation}
to describe the results obtained by numeric simulation.

Probability $p(G)$ is shown in Fig. \ref{mudry_hist.ps} as continuous lines.
Although the three numerical distributions are nicely reproduced,
accuracy is better for wider wires as expected from the theory
($M \ll L$ is assumed in the theory leading to Eq. (\ref{p_law})). 
The divergence at $G=1$ is of the inverse square root form
while the apparent $G^{-1}$ non--integrable divergence at the origin
is regularized by the complex numerator which vanishes at $G=0$.
In conclusion, Eq.(\ref{p_law}) is a very good description of the
numerical data once the constant $C$ is properly estimated.

\begin{figure}
\epsfig{figure=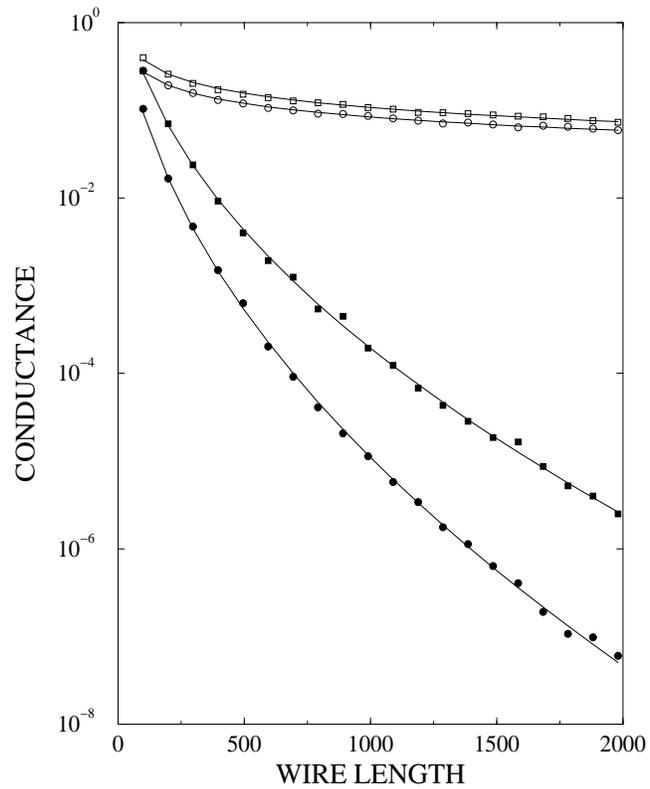,height=12.6cm,width=9cm}
\caption{Length scaling of the conductance of a wire of 9 (circles)
or 19 channels (squares) at the band center ($E=0$).
Scaling of the median (filled symbols) is shown besides the data of
Fig. \ref{mudry1.ps} in a semilogarithmic plot.
Medians can be fitted over the whole range of studied lengths by
(\ref{law2}) giving values of $\xi$ equal to 5.73 and 8.99
for the 9 and 19 channels wires, respectively.}
\label{mudry2.ps}
\end{figure}

Let us try an alternative way of characterizing the central value of
distributions of the form shown in Fig. \ref{mudry_hist.ps}.
Previous experience with disorder models of this kind
proves that the median is more robust than the mean
for some distributions showing large $G$ variances but smaller $\ln (G)$
variances, i.e., when geometric mean does it better that the usual arithmetic
mean\cite{jav2}. Unfortunately, the distribution of $\ln (G)$ is also
very broad in the present case.
Scaling results for the toy model of Section II suggest the use of the median.
Fig. \ref{mudry2.ps} gives the median scaling
obtained for the previously collected conductance distributions.
In can be seen that median values are
exponentially smaller than mean values at large wire lengths.
The fact that the median scales towards 0 can be inferred from the
scaling of the histograms given in Fig. \ref{mudry_hist.ps}
although, the accumulation near
exponentially small values of $G$ is not visible in the Figure.
Even the scaling law of medians is the same that describes the chain
model (see, Eq. (\ref{law2})).
Alternatively, the exponential decay of the median as a function of
the square root of the wire length can be inferred from 
Eq. (\ref{p_law}) once the distribution is written as a logarithmic
normal distribution using the variable change described in the
next subsection.

The practical meaning of the result is quite clear, more than half
of the conductance observations are exponentially decreasing as the
wire length increases. Actually, the fraction of the samples showing
exponentially small conductances increases with wire length because probability
below $G=1$ decreases monotonically. Just the difference with a conventional
one--dimensional scaling (as, for example, the one obtained for wires of
even width) comes from the power law decrease of the upper part of the
probability distribution that should be compared with the exponential
decrease characteristic of standard scaling.

\subsection{Two dimensional system}

\begin{figure}
\epsfig{figure=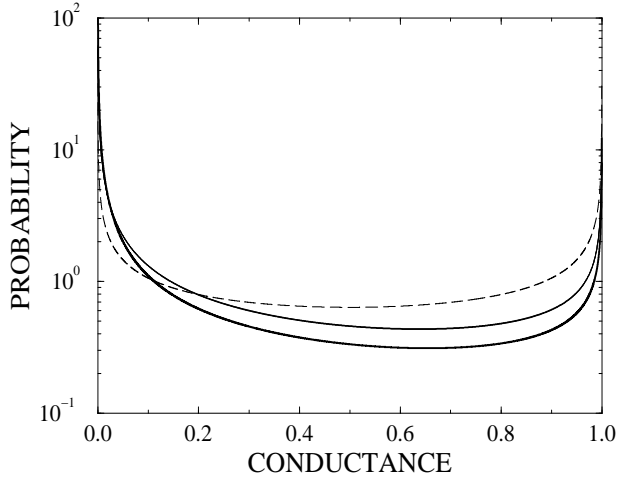,height=7cm,width=9cm}
\caption{Probability distribution of conductances between point contacts
in a large disordered square sample.
Results for three sizes ($L=599, 699, 799$) are given
together with the theoretical prediction given by Eq. (\ref{p_law})
with $C=0.0948$ (All four lines are seen as the thin line in the figure).
The {\it universal} distribution
corresponding to the infinite two--dimensional system is given by
the thick line ($C=0.0471$). The conductance distribution corresponding
to the Porter-Thomas wavefunction statistics is given by the
dashed line. Conductance unit is ${e^2} / h$.}
\label{g0_dist.ps}
\end{figure}

While a standard study of the scaling of the conductance in 2D would imply
the calculation of conductances of increasing $L \times L$ samples connected
to ideal leads of width equal to the square side $L$, in this work
I have used a dot setup specially designed for the study of just one
conducting state\cite{note1}.
Certainly, when the number of incoming channels is fixed
by a point contact geometry, the only factor that affects the value of the
conductance is the size (area) of the dot. In this way, the presumably
increase of the conductance due to wider contacts does not obscure the
underlying scaling law strictly due to the increased size of the system.
Two different limits are well known for large values of $L$. First, ballistic
transport through the sample can occur as it happens in chaotic cavities.
This limit is described in a first approximation by scattering matrices of the
Circular Orthogonal Ensemble (COE)\cite{coe}. Roughly speaking, a conductance
about $1 \over 2$ per channel can be expected. Second, Anderson exponential
localization would imply an exponentially small value of conductances for
large enough dot sizes. This limit applies to diagonal disorder, for example.
At the band center of a system with chiral symmetry, numerical
simulation shows a behavior similar to the ballistic one
(see Fig. \ref{g0_dist.ps}). Mean conductance
converges to a well--defined finite limit while the whole conductance
distribution is {\it perfectly} described by Eq. (\ref{p_law}). Since
now the system is not quasi--onedimensional as wide wires are, one is
forced to conclude that there should be deep general reasons for the
validity of Eq. (\ref{p_law}) in this context.

Let me briefly describe the numerical procedure. The transmission between
any couple of points within the dot is calculated
and the corresponding conductance distribution obtained.
To this end, two clean infinite chains (the leads) are attached through
two arbitrarily chosen lattice sites within the disordered square sample
(the dot). For this geometry, the transmission from site $\bf r$
to site $\bf r'$ is given by the following expression:

\begin{equation}
T={{4\ {\psi (\bf r)}^2\ {\psi (\bf r')}^2} \over
\big( {\psi (\bf r)}^2 + {\psi (\bf r')}^2 \big)^2}  \qquad,
\label{transmission}
\end{equation}

\noindent
where $\psi(\bf r)$ is the wavefunction at the band center.
From this equation, $0 \le T \le 1$, and the conductance is between
zero and one quantum unit for this numerical simulation.
While the large number of transmission evaluations ($\sim \frac{1}{2} L^4$)
allows for a very precise calculation of the whole conductance distribution,
the dependence of the conductance on the separation between point contacts
is not given by the procedure.
The eigenstate at the band center is obtained by direct inversion of the
Schr\"odinger equation ($\hat H \psi =0$) where $\hat H$ is given by
(\ref{RHS}).
Once the $E=0$ wavefunction statistics is known, Eq. (\ref{transmission})
can be used to get a conductance distribution.
For example, if the Porter-Thomas form\cite{poto} were valid:

\begin{equation}
f(t)={1 \over {\sqrt{2 \pi t}}} \exp(-t/2) \qquad,
\end{equation}

\noindent
where $t=N {\psi (\bf r)}^2$ being $N$ the number of sites,
the conductance distribution would be given by:

\begin{equation}
p(G)=
\int_0^\infty \int_0^\infty {dt\ dt'\ f(t)\ f(t')\
\delta\Big(G-{{4 t t'} \over (t + t')^2}}\Big) \qquad.
\end{equation}

\noindent
which can be integrated to give the final result:

\begin{equation}
p_{\rm PT}(G)={1 \over \pi}{1 \over \sqrt{G (1-G)}} \qquad.
\label{pt_law}
\end{equation}

Although probability $p_{\rm PT}(G)$ reproduces some features of the
conductance distributions
shown in Figs. \ref{mudry_hist.ps} and \ref{g0_dist.ps}
for the smaller cluster sizes
(for example, the square root divergence at $G=1$),
it is clearly non comparable
to the accurate result given by Eq.(\ref{p_law}).
What comes as a surprise is the fact that a logarithmic normal
distribution of the wavefunction ratios $t/t'$ exactly gives
the conductance distribution proposed by Mudry {\it et al.}\cite{mudry}.
That is, assuming

\begin{equation}
g(x)=\sqrt{\frac{C}{4 \pi}} \exp(- \frac{C}{4} x^2)
\label{lognormal_law}
\end{equation}

\noindent
being $x=\ln(t/t')$, the conductance distribution is given by

\begin{equation}
\int_{-\infty}^\infty  {dx\  g(x)\ 
\delta\Big(G-{\frac{4}{e^x+e^{-x}+2}}\Big) }\qquad.
\end{equation}

\noindent
This integral is easily solved giving $p(G)$ of Eq.(\ref{p_law}) with

$$
G=\frac{4}{e^x+e^{-x}+2}
$$

\noindent
or equivalently,

$$
\frac{x}{2} = {\rm arcosh}(G^{-1/2}) \qquad.
$$

\begin{figure}
\epsfig{figure=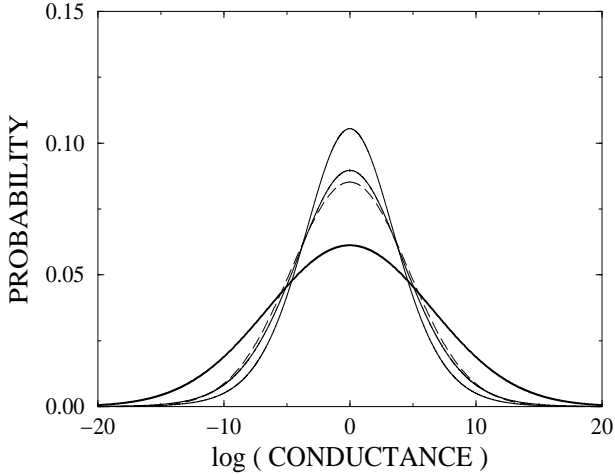,height=7cm,width=9cm}
\caption{Probability distribution of the wavefunction weight ratio
in a square disordered dot. Results for two sizes ($L=99$ and $499$)
are given as continuous lines together with the theoretical prediction
(Eq. (\ref{lognormal_law})) for $C=0.09124$ which corresponds to $L=499$
(dashed line). The {\it universal} distribution
corresponding to the infinite two--dimensional system is given by
the thick line ($C=0.0471$).}
\label{lognormal.ps}
\end{figure}

\noindent
This result gives some clue over the complicate $G$ dependence that
happens to appear in its distribution function.
Numerical simulation (see Fig. \ref{lognormal.ps})
shows that Eq.(\ref{lognormal_law}) accurately
describes the wavefunction squared ratios of large disordered
two-dimensional systems and, consequently,
conductance distributions of the form given by Eq.(\ref{p_law})
are valid in this case.

\begin{figure}
\epsfig{figure=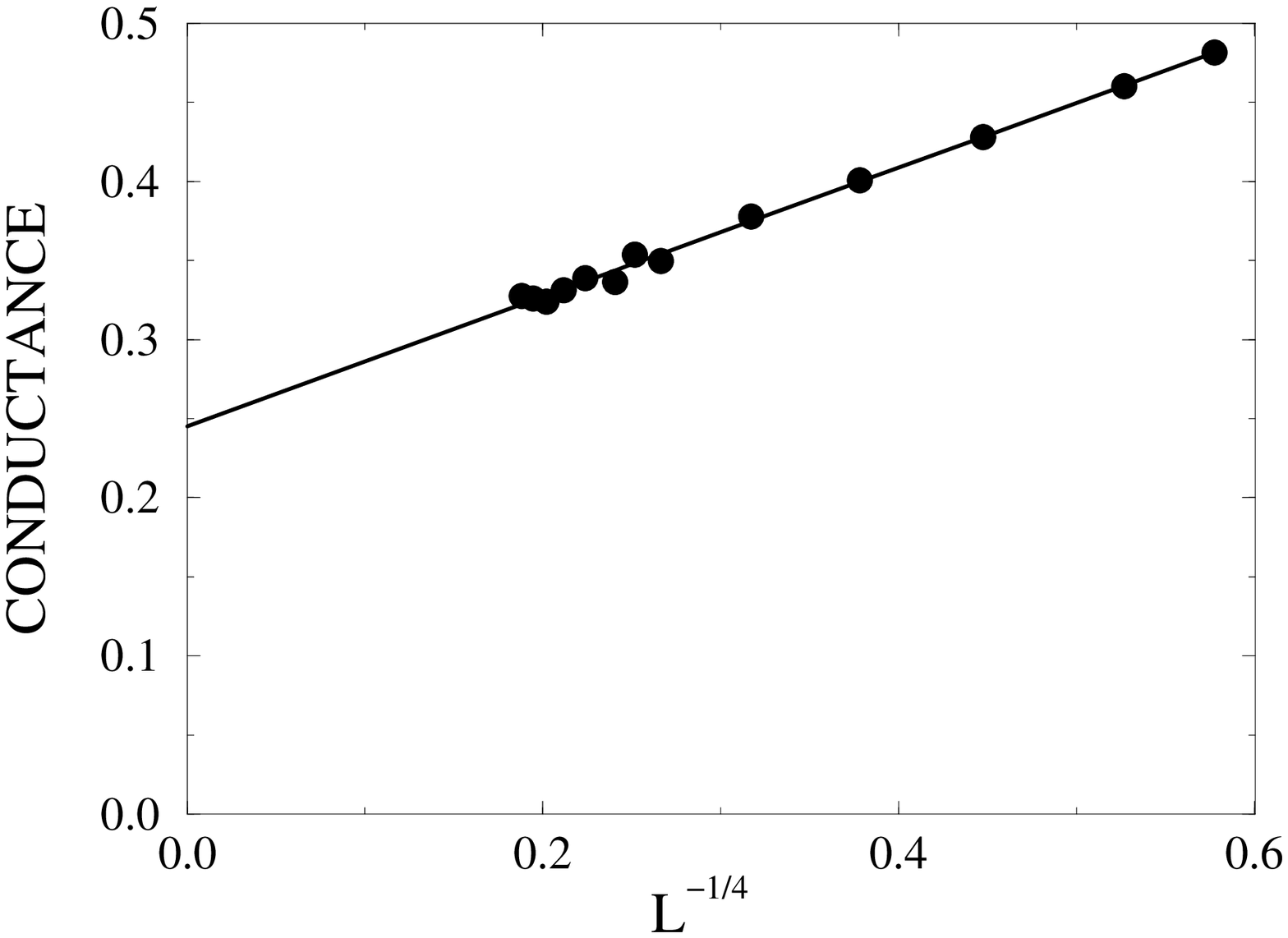,height=7cm,width=9cm}
\caption{Size scaling of the mean value of the conductance
between point contacts in a square disordered dot.
Continuous line is a fit of the form given by Eq.(\ref{2d_law})
with an asymptotic value of $a=0.245$.
Conductance is given in units of the conductance quantum ${e^2} / h$.}
\label{g0.ps}
\end{figure}

Let us now discuss the scaling properties of the mean conductance
at the band center. Conductance has been averaged over
a significant number of randomly generated samples of increasing
linear sizes ($L$) keeping almost constant the total number of
{\it measurements}.
Fig. \ref{g0.ps} shows the results obtained by this numerical procedure.
The scaling of the mean conductance can be fitted to a model of the form:

\begin{equation}
<G>=a + b L^{-\frac{1}{4}}
\label{2d_law}
\end{equation}

\noindent
with a very good precision. The asymptotic value corresponding to
the infinite limit is 0.245. Finite values of the conductance of
two--dimensional systems showing chiral symmetry have been predicted in a
number of papers\cite{gade,furusaki,mudry0,campos,mudry2}.
Although, the present numerical
simulation nicely supports these theories some caution must be used for
two main reasons. First, a somewhat practical reason.
Results heavily depend on the fact that the
Green function of this problem is just given by only one particular
state. While the energy of this state is well defined theoretically,
it could be difficult to make an experiment at just a particular energy.
Previous authors on the subject have clearly shown that chirality is lost
as soon as $E=0$ is left\cite{mudry}.
Second reason is a bit more technical. Scaling properties have been obtained
for clusters of an odd number of sites and, therefore, a state at $E=0$.
Present computational facilities do not allow to prove that scaling of
clusters of an even number of states proceeds in the same way. (Note that
in this case the Green function should be recalculated for any new position
of the point contacts since storing of the whole Green function matrix of
the isolated cluster is not possible). Nevertheless, I have checked
that odd/even differences are minimal for small systems.

The {\it universal} conductance distribution at the band center of a
two--dimensional chiral system shown by the thick line in
Fig.(\ref{g0_dist.ps}) is probably the most original result in the paper.
It corresponds to the $L\to\infty$ limit of Eq.(\ref{2d_law})
which is obtained for $C=0.0471$ in Eq.(\ref{p_law})
but does not differ very much from the distributions
corresponding to the simulated sizes ($L=599,699,799$) plotted as
three thin indistinguishable lines. All three
are accurately described by Eq.(\ref{p_law}) with $C=0.0948$.
Although obtained for point contacts,
previous experience with wide wires proves that minor differences
should be expected for wider but finite contacts\cite{note2}.
Let me insist in the idea
that perfect agreement with the theoretical conductance distribution
is also explained by a particular distribution of wavefunction
weight ratios (see Eq.(\ref{lognormal_law})) that is confirmed by
numerical simulation. Fig. \ref{lognormal.ps} give the distributions
obtained for a small cluster size ($L=99$) -higher maximum value-
and for a larger size ($L=499$) together with its corresponding
Gaussian (dashed line). The thick line gives the {\it universal} prediction
for this statistical measure of the wavefunction.
Consequently, I think that
unusual phenomena as they have been described
in the literature are quite compatible with present results for the
conductance distribution\cite{furusaki,mudry2,campos}.

\section{Discussion}

The scaling properties of the conductance distribution of wide wires
showing chiral symmetry has been analyzed by numerical solution
of a simple model Hamiltonian. Predictions of
a power law decrease of the mean conductance have been confirmed.
Besides, the strange scaling properties of the distribution give rise
to unusual statistical properties that have been described in the
main text. For example, while the slow algebraic decay
of the mean conductance is due to few but very large fluctuations,
it has been shown that almost all conductance measurements
are exponentially small for large wires.
Actually, an exponential law is obtained for the median of the distribution
and for any generalization of the median based on the integral
of the distribution. Nevertheless, exponential decay is not proportional
to the sample length but to its square root.
That is a significant deviation from standard localization theory.

Surprisingly, statistical conductance properties of two--dimensional
systems connected to point contacts are described at the band center
by the same distribution that describes wide wires.
It has been shown that this statistics appears due to the existence of
an underlying Gaussian distribution of the logarithm of the ratio
of the wavefunction weights at $E=0$.
The main difference with quasi one--dimensional systems is that now
both the mean and the median conductance seem to reach a finite limit.
The {\it universal} statistical distributions of conductance and wavefunction
ratios have been obtained. Although they share some properties with
the corresponding statistics of the Circular Orthogonal Ensemble,
it has been shown that they differ.
This numerical result calls for some theory able to explain the origin
of such a simple statistics of the wavefunction at the band center of a
chiral system.

\acknowledgments
This work has been partially supported by Spanish 
Comisi\'on Interministerial de Ciencia y Tecnolog\'{\i}a (grant PB96-0085)
and Direcci\'on General de Ense\~nanza Superior e Investigaci\'on y Ciencia
(grant 1FD97-1358).


\newpage

\begin{thebibliography}{}
\bibitem{curiosidad}
If the complex probability distribution of selfenergies
implied by Eq.(\ref{chain4}) were substituted by its geometric mean,
a value of 1 would be obtained. But this does not imply
that perfect transmission ($G=1$) or
diverging localization length in the language of other authors is the
average value of the relevant transport magnitude.
Actually, both the distribution of selfenergies or Lyapunov exponents
are broad and cannot be substituted by any central value of their
distributions. Ref. (\onlinecite{altshuler}) further illustrates this point.
\bibitem{note2}
Situation quantitatively changes when
contacts of the order of the square side are considered\cite{note1}.
In this case, the conductance average is larger ($\sim 1.74$) and its
distribution Gaussian.
\end{thebibliography}
\end{document}